\documentclass[aps,pra,twocolumn]{revtex4}
\newcommand {\be}{\begin{equation}}
\newcommand {\ee}{\end{equation}}
\newcommand {\bey}{\begin{eqnarray}}
\newcommand {\eey}{\end{eqnarray}}
\usepackage{epsfig}
\usepackage{amsfonts}
\usepackage{amsmath}
\usepackage{mathbbol}

\begin{document}

\title{Approximate simulation of entanglement with a linear cost of communication}

\author{A. Montina}
\affiliation{Perimeter Institute for Theoretical Physics, 31 Caroline Street North, Waterloo, 
Ontario, Canada N2L 2Y5}

\date{\today}

\begin{abstract}
Bell's theorem implies that the outcomes of local measurements on two maximally entangled 
systems cannot be simulated without classical communication between the parties. The
communication cost is finite for $n$ Bell states, but it grows exponentially in $n$.
Three simple protocols are presented that provide approximate simulations for low-dimensional 
entangled systems and require a linearly growing amount of communication. 
We have tested them by performing some simulations for a family of measurements.
The maximal error is less than $1\%$ in three dimensions and grows sublinearly with 
the number of entangled bits in the range numerically tested. One protocol is the 
multidimensional generalization of the exact Toner-Bacon 
[Phys. Rev. Lett. {\bf 91}, 187904 (2003)] model for a single Bell state.
The other two protocols are generalizations of an alternative exact model, which we derive 
from the Kochen-Specker [J. Math. Mech. {\bf 17}, 59 (1967)]
scheme for simulating single-qubit measurements. These
protocols can give some indication for finding optimal one-way communication protocols 
that classically simulate entanglement and quantum channels. Furthermore they can be 
useful for deciding if a quantum communication protocol provides an advantage on classical 
protocols.
\end{abstract}
\maketitle

\section{Introduction}

For some tasks in information processing, quantum communication channels were proved
to be much more powerful than classical channels. While Holevo's theorem~\cite{holevo} 
states that $n$ qubits cannot encode a message of more than $n$ classical bits, quantum 
channels reveal their real power when separate devices need to exchange information to 
jointly perform a task whose result depends on the data held by any single party. 
In distributed computing, the communication complexity of a problem is the cost in 
communication of the most efficient solution for that problem~\cite{nisan}. 
Similarly, the quantum communication complexity is the minimal amount of quantum 
communication required to accomplish a distributed computation. Quantum protocols for 
some communication problems, such as the Raz's problem~\cite{raz} or the hidden matching 
({\it HM}) problem~\cite{baryossef}, made clear that in some cases quantum channels can 
be exponentially more powerful than classical channels. Indeed, the {\it HM} 
problem exhibits an $n$ (qubits) versus $2^{\Omega(\sqrt{n})}$ (bits) gap between the 
quantum and classical communication complexity for bounded-error protocols. 
In the case of the Deutsch-Jozsa problem and errorless protocols, the gap is even stronger:
$n$ qubits versus $\Omega(2^n)$ bits~\cite{buhrman0}.
A review on quantum communication complexity can be found in Ref.~\cite{buhrman}.

A protocol of communication complexity is supposed to produce a correct result at least 
with high probability. Let us consider the scenario introduced by Yao~\cite{yao}. Two 
parties, say Alice and Bob, get a fixed part of the input data. Alice's input is an 
element $m_A$ of a finite set $M_A$ and Bob's input is an element $m_B$ of a finite 
set $M_B$. The purpose is computing a function $f(m_A,m_B)$ with a probability of success 
close to $1$ by some amount of communication between the parties. In a broader scenario 
one can just require that the outcome is generated according to a given probability 
distribution depending on $m_A$ and $m_B$. The problem of classically simulating 
a quantum communication channel fits into this extended framework. A quantum
state is prepared by Alice through a procedure $m_A$ and sent to Bob through a quantum 
channel. He eventually performs a measurement through a procedure $m_B$. The measurement 
outcome, say $\beta$, is generated with a probability, $\rho(\beta|m_A,m_B)$,
that depends on both $m_A$ and $m_B$. The task of
a classical simulation is reproducing the distribution $\rho$ by replacing the 
quantum communication with a classical communication. This task is a particular 
problem of communication complexity in the outlined generalized sense. We call 
the minimal amount of classical communication needed in the simulation 
{\it classical communication complexity of the quantum channel} (the classical 
communication is not supposed to be necessarily one-way). Of course, such a 
simulation requires at least the same communication resources that are necessary for 
classically simulating a quantum communication protocol with an (almost) deterministic 
outcome. Thus, the previously mentioned results in quantum communication complexity 
sharpen the conceptual differences between quantum and classical physics. 
Indeed, they imply that any classical description of $n$ qubits 
requires a quantity of resources growing at least as $O(2^n)$~\cite{brassard} 
(or $2^{O(\sqrt n)}$ if a bounded error is admitted), a property 
that hardly fits into the framework of classical physics, where in general the 
quantity of resources scales linearly with the 
amount of information virtually accessible in an experiment. Thus, a peculiar 
feature of quantum physics is the necessity of a huge quantity of information that 
is almost completely concealed from a direct experimental observation, but it is 
nevertheless fundamental in the description of systems. Indeed the whole information 
encoded in the quantum state cannot be used for carrying information~\cite{holevo}, 
but most of that is fundamental in the description of some quantum protocols of
communication complexity. 

There is an open question concerning the classical communication complexity of a quantum 
channel. On the one hand, the best known protocol for simulating a quantum channel uses 
an amount of resources that scales as $n 2^n$~\cite{buhrman,massar}, even with a bounded 
error. On the other hand, the {\it HM} problem gives the lower bound $2^{\Omega(\sqrt{n})}$ 
for the minimal amount of communication in the case of bounded error. At present, no other 
constraint is known; thus one could hope to find a better bounded-error simulation of a 
quantum channel with communication complexity scaling as $2^{\sqrt n}$. 

As established by Bell's theorem~\cite{bell}, correlations of outcomes produced 
by local measurements on entangled systems cannot be explained classically
without post-measurement communication between the parties.
The problem of quantifying the classical communication complexity of a quantum
channel of $n$ qubits is essentially equivalent to the problem of finding the 
minimal amount of communication needed to classically simulate the outcomes
of local measurements on $n$ Bell states. Indeed 
classical protocols for simulating quantum channels can be converted into
classical models of entanglement without affecting the cost of communication.
Conversely, a classical model of entanglement can be converted into
a model of quantum channels with a little more communication, as shown in
Ref.~\cite{cerf} in the case of a single Bell state. A more general proof will be
given in Sec.~\ref{sec2B}, where we will show that an increase
of communication by $n$ bits on average is sufficient for the conversion.

In this paper, we present three approximate classical protocols for simulating 
bipartite entanglement that need a one-way communication equal to 
the number of entangled bits (ebits). One of these is a multidimensional 
generalization of an exact model for a single ebit, reported by Toner and Bacon in 
Ref.~\cite{toner}. The other two protocols are generalizations of an alternative 
exact model, which will be derived here.
The accuracy of the protocols was numerically tested for a family of measurements
and for a
dimension of the Hilbert space (of each party) between $2$ (one ebit) and $32$ 
($5$ ebits). The maximum error is less than $1\%$ in the three-dimensional case 
and increases sublinearly with the number of ebits in the tested range (Note 
that the error cannot be bounded, since, as previously said, a protocol with
bounded error needs at least $2^{O(\sqrt{n})}$ bits of communication). Our 
models can be useful for two reasons. First, they can give an indication for
finding an optimal algorithm with bounded error, as discussed in the
concluding remarks of the paper. Second, even if the
maximal discrepancy increases with the number of qubits, these models or some
modified version can give accurate results for measurements and
states involved in some protocols of quantum communication complexity. 
In such a case they would give a proof that these quantum protocols do not
provide any advantage on classical protocols.

In Sec.~\ref{sec1}, we review the Toner-Bacon protocol for a single Bell state 
and derive the alternative exact protocol starting from the Kochen-Specker
hidden variable model of a qubit~\cite{kochen}. The three generalized 
protocols for higher dimensions of the Hilbert space and their Monte Carlo 
simulations are presented in Sec.~\ref{sec2A}. In Sec.~\ref{sec2B} a
general method for converting an entanglement model into a quantum
channel model is presented. Finally, we draw the
conclusions and perspectives in the last section.

\section{Simulation of Bell states with one bit of communication}
\label{sec1}

Two qubits are prepared in the Bell state 
\be\label{singlet}
|\Psi_s\rangle=\frac{1}{\sqrt2}\left(|1\rangle|2\rangle-
|2\rangle|1\rangle\right)
\ee
and each one is sent to two parties, 
Alice and Bob, who perform a local measurement on their own qubit. Bell's 
theorem~\cite{bell} implies that the outcome probabilities cannot be explained 
by a local classical theory and any exact simulation of this scenario needs some 
communication between the parties. How much information has to be 
exchanged? Trivially Bob could send the whole information about the measurement 
he performed, which is actually infinite. However, Brassard et al.~\cite{brassard} 
showed that a finite amount of communication is sufficient for exactly reproducing 
the outcome probabilities. They presented a model that simulates Bell correlations 
and requires exactly $8$ bits of communication. 
Steiner~\cite{steiner} reported a different model, which requires $2.97$ bits on 
average, but the amount of communication for each particular realization is unbounded. 
This result was improved in Ref.~\cite{cerf}, where the average information was 
lowered to $2.19$ bits. Later, Toner and Bacon 
showed that just one bit of communication can account for Bell correlations~\cite{toner}
and two bits are sufficient for simulating teleportation of a qubit. In the next 
subsection we review this model. In Sec.~\ref{mymodel} an alternative exact model 
is derived whose direct generalization to $N$ dimensions is one of the
three protocols described in Sec.~\ref{sec2A}.

\subsection{Toner-Bacon model}
\label{tonerSec}

Let us consider the previously described scenario with two qubits in the Bell
state~(\ref{singlet}). Alice and Bob perform two projective measurements. 
Bob's measurement projects the quantum state into one of two mutually orthogonal 
vectors of the two-dimensional Hilbert space. Let us represent them by the Bloch 
vectors $\vec b_1$ and $\vec b_2=-\vec b_1$. 
Alice performs a projective measurement on the states $\vec a_1$ and $\vec a_2=-\vec a_1$.
The joint probability of having outcomes $\vec a_\alpha$ and $\vec b_\beta$ is
\be\label{quant_prob}
P(\alpha,\beta)=\frac{1}{4}\left(1-\vec a_\alpha\cdot\vec b_\beta\right).
\ee
It is possible to exactly reproduce this statistics through a classical 
protocol that uses just one bit of classical communication. The protocol
is as follows. Alice and Bob share two random unit vectors $\vec\lambda_1$
and $\vec\lambda_2$. They are uncorrelated and uniformly distributed on
the unit sphere. Bob generates the outcome $\beta$ such that the vector 
$\vec b_\beta\in\{\vec b_1,\vec b_2\}$ is closest to $\vec\lambda_1$, that is, 
\be\label{rule_bob}
\mathrm{sgn}(\vec b_\beta\cdot\vec\lambda_1)>0.
\ee
Bob sends one bit $n\in\{-1,1\}$ to
Alice, where 
\be\label{rule_comm}
n=\mathrm{sgn}(\vec b_\beta\cdot\vec\lambda_2).
\ee
Alice generates the outcome $\alpha$ such that 
\be\label{rule_alice}
\mathrm{sgn}\left[\vec a_\alpha\cdot(\vec\lambda_1+n\vec\lambda_2)\right]<0.
\ee
This protocol produces the events according to the quantum probability 
$P(\alpha,\beta)$, as proved in Ref.~\cite{toner}.
Note that the effect of $n$ is to change the sign of $\vec\lambda_2$, that is, Alice 
receives the instruction ``if $n$ is negative, flip the vector $\vec\lambda_2$ to
the opposite direction''. 

It is useful to express the protocol in a form that is trivial to generalize to higher 
dimensions of the Hilbert space. For this purpose we replace the Bloch vectors
with vectors in the two-dimensional Hilbert space. While the state~(\ref{singlet})
simplifies the notation with Bloch vectors, for a generalization to higher dimensions
it is better to use the state
\be\label{bstate}
|\Psi\rangle=\frac{1}{\sqrt2}\left(|1\rangle|1\rangle+|2\rangle|2\rangle\right).
\ee
The two states~(\ref{singlet},\ref{bstate}) differ by the local unitary transformation
$e^{i\frac{\pi}{2}\hat\sigma_y^{(2)}}$, performed on the second qubit.

The projective measurements are represented by two sets of orthogonal vectors, 
\be
\{|\phi_1\rangle,|\phi_2\rangle\}\equiv M_A, \;\;
\{|\psi_1\rangle,|\psi_2\rangle\}\equiv M_B
\ee
Alice and Bob measure $M_A$ and $M_B$, respectively. Each vector in the set is associated
with one outcome. In the simulation, the shared vector $\vec\lambda_1$ is replaced by a vector
$|x\rangle$ in the two-dimensional Hilbert space,
\be
\vec\lambda_1\rightarrow|x\rangle,
\ee
so that $\vec\lambda_1$ is the Bloch vector of $|x\rangle$. The vector 
$\vec\lambda_2$ is replaced by a set of orthogonal vectors in the Hilbert space, 
\be
\vec\lambda_2\rightarrow\{|y_1\rangle,|y_2\rangle\},
\ee
so that
$\vec\lambda_2$ and $-\vec\lambda_2$ are the Bloch vectors of $|y_1\rangle$
and $|y_2\rangle$, respectively. In this new frame, the Toner-Bacon protocol
is as follows. According to Eq.~(\ref{rule_bob}), Bob generates the
outcome $|\psi_\beta\rangle\in M_B$ that is closest to $|x\rangle$ (he
maximizes $|\langle\psi_\beta|x\rangle|^2$). Then he 
sends Alice the index $n\in\{1,2\}$ such that $|y_n\rangle\in 
\{|y_1\rangle,|y_2\rangle\}$ is the vector closest to $|\psi_\beta\rangle$
[from Eq.~(\ref{rule_comm})]. Let us introduce the vector
\be
|\phi_k^*\rangle\equiv \langle\phi_k|1\rangle|1\rangle+\langle\phi_k|2\rangle|2\rangle
\ee
for $k=1,2$.
It is obtained from $|\phi_k\rangle$ by the complex conjugation of the
coefficients in the basis $\{|1\rangle,|2\rangle\}$. Alice generates the 
outcome $|\phi_\alpha\rangle\in M_A$ that maximizes the function 
$$\langle\phi_\alpha^*|\left\{|x\rangle\langle x|+|y_n\rangle\langle y_n|\right\}
|\phi_\alpha^*\rangle.$$
Note that Eq.~(\ref{rule_alice}) would ask to minimize
$\langle\phi_\alpha|\left\{|x\rangle\langle x|+|y_n\rangle\langle y_n|\right\}
|\phi_\alpha\rangle$ in the case of 
state (\ref{singlet}). It is easy to show that for state~(\ref{bstate}) 
``minimization'' is replaced by ``maximization'' and 
$|\phi_\alpha\rangle$ by $|\phi_\alpha^*\rangle$.
The outcomes are generated according to Born's rule, that is,
\be
P(\alpha,\beta|M_A,M_B)=\frac{1}{2}|\langle\phi_\alpha|\psi_\beta\rangle|^2.
\ee

This new reformulation of the Toner-Bacon protocol suggests a very simple
generalization to higher dimensions of the Hilbert space, as
discussed in Sec.~\ref{sec2A}.

\subsection{Alternative model}
\label{mymodel}

An alternative exact classical model for a Bell state can be obtained from the hidden 
variable model of a 
qubit introduced by Kochen and Specker (KS)\cite{kochen}. This can be seen as 
a classical model of a quantum channel where the communicated classical information
is infinite and encoded in a three-dimensional unit vector. It can be
transformed into a model with finite communication cost, as shown below.

\subsubsection{Quantum channel}
\label{qchann_sec}

Let us introduce the KS model. It provides a classical simulation of
the following scenario. Bob prepares a qubit in a quantum state
represented by a Bloch vector $\vec b$. He sends the qubit to
Alice, who performs a projective measurement $M_A=\{\vec a_1,\vec a_2\}$
with outcome states $\vec a_1$ and $\vec a_2\equiv-\vec a_1$.
The classical simulation is as follows.
Given the Bloch vector $\vec b$, Bob generates a unit vector 
$\vec x_1$ with probability
\be\label{prob_ks}
\rho(\vec x_1|\vec b)=\frac{1}{\pi}\vec b\cdot\vec x_1
\theta(\vec b\cdot\vec x_1),
\ee
where $\theta(z)$ is the Heaviside function, which is equal to $1$
if $z>0$ and zero otherwise. He sends $\vec x_1$ to Alice. Given 
the vector pair $M_A=\{\vec a_1,\vec a_2\}$,
Alice generates the outcome $\vec a_\alpha$ according to the conditional 
probability 
\be\label{cond_prob}
P(\alpha|\vec x_1;M_A)=\theta(\vec a_\alpha\cdot\vec x_1);
\ee
in other words, she generates deterministically the vector closest to $\vec x_1$.
This model gives the correct quantum probability for the
outcomes, that is,
\be\label{quant_equiv}
\int d^2 x_1 P(\alpha|\vec x_1;M_A)\rho(\vec x_1|\vec b)=
\frac{1}{2}\left(1+\vec a_\alpha\cdot\vec b\right).
\ee

Let us define the vectors
\be\begin{array}{l}
\label{vect_y}
\vec y_1\equiv\vec x_1+\vec x_2, \\
\vec y_2\equiv\vec x_1-\vec x_2. 
\end{array}
\ee
It is easy to show that the probability distribution~(\ref{prob_ks}) 
can be obtained from the uniform distribution
\be\label{uni_prob}
\rho(\vec x_1,\vec x_2|\vec b)=\frac{1}{4\pi^2}
\theta(\vec b\cdot\vec y_1)\theta(\vec b\cdot\vec y_2)
\ee
by integrating out the unit vector $\vec x_2$. The integration
is trivial in spherical coordinates and is left as an exercise.
This uniform probability distribution and the conditional
probability~(\ref{cond_prob}) still give the correct quantum
predictions. More generally we get an exact simulation of
a qubit with the probability distributions
\bey\label{mod_n1}
\rho(\vec x_1,\vec x_2|n_1,n_2;\vec b)=\frac{1}{4\pi^2}
\theta\left(n_1\vec b\cdot\vec y_1\right)   
\theta\left(n_2\vec b\cdot\vec y_2\right)    \\
\label{cond2}
P(\alpha|\vec x_1,\vec x_2,n_1,n_2;M_A)=\theta\left[\vec a_\alpha\cdot(n_1 
\vec y_1+n_2\vec y_2)\right]
\eey
for any $n_i=\pm1$. For example, with $n_1=1$ and $n_2=-1$, Eqs.~(\ref{mod_n1},
\ref{cond2}) can be derived from Eqs.~(\ref{cond_prob},\ref{uni_prob})
by exchanging $\vec x_1$ and $\vec x_2$. With $n_1=-1$ and $n_2=-1$ we have to 
flip the direction of $\vec x_1$ and $\vec x_2$. Finally, with $n_1=-1$ and
$n_2=1$ we have to exchange and flip the vectors.

Suppose that the indices $n_i$
are randomly generated with probability $\rho_I(n_1,n_2)\equiv 1/4$.
The joint probability distribution of $\vec x_i$ and $n_i$ can
be suitably written in the form
\be\label{distr2}
\rho(\vec x_1,\vec x_2,n_1,n_2|\vec b)=\rho(n_1,n_2|\vec x_1,\vec x_2;\vec b)
\rho_v(\vec x_1,\vec x_2),
\ee
where
\be\label{prob_n}
\rho(n_1,n_2|\vec x_1,\vec x_2;\vec b)\equiv
\theta\left(n_1\vec b\cdot\vec y_1\right)\theta\left(n_2\vec b\cdot\vec y_2\right)
\ee
and
\be\label{distr_x}
\rho_v(\vec x_1,\vec x_2)\equiv\frac{1}{(4\pi)^2}.
\ee
The model with probability distribution~(\ref{distr2}) and conditional 
probability~(\ref{cond2}) gives again the correct quantum probabilities, but now 
we have the nice property that the marginal probability distribution 
$\rho_v(\vec x_1,\vec x_2)$
of the vectors $\vec x_1$ and $\vec x_2$ does not depend on the prepared 
quantum state $\vec b$. This dependence is only in the conditional probability 
distribution $\rho(n_1,n_2|\vec x_1,\vec x_2;\vec b)$ of the discrete indices
$n_i$.
This reformulation of the Kochen-Specker model gives a protocol with
shared noise for simulating the communication of a qubit with just two 
bits of classical communication: Bob and Alice share the random
vectors $\vec x_1$ and $\vec x_2$, generated according to the
uniform probability distribution~(\ref{distr_x}); given a quantum
state $\vec b$, Bob generates the discrete indices $n_i$
according to rule~(\ref{prob_n}); he sends them to Alice; given
a measurement $M_A$, Alice generates the outcome $\vec a_\alpha$
according to rule~(\ref{cond2}).

\subsubsection{Entanglement}

Any classical simulation of a quantum channel can be 
converted into a simulation of entanglement without increasing the amount of
communication. Consider again the scenario of Sec.~\ref{tonerSec} with two qubits 
in the Bell state~(\ref{singlet}). Alice and Bob perform two projective measurements, 
$M_A=\{\vec a_1,\vec a_2\}$ and $M_B=\{\vec b_1,\vec b_2\}$, respectively, and
get outcomes $\vec a_\alpha$ and $\vec b_\beta$. The marginal probability of 
$\beta$ is uniformly distributed on the values $1$ and $2$. Furthermore,
given Bob's outcome $\vec b_\beta$, Alice's outcome is generated as if she 
received the quantum state $-\vec b_\beta$ directly from Bob. Thus, the joint 
probability distribution of the outcomes can be simulated as follows. Bob randomly 
generates the outcome $\vec b_\beta$ with uniform probability distribution 
$\rho(\beta)=1/2$. He then uses a classical model of a quantum channel for
sending the quantum state $-\vec b_\beta$ to Alice, who finally generates
her own outcome $\vec a_\alpha$. The amount of communication of the derived 
entanglement model is equal to that of the quantum channel model.

The classical model of a quantum channel previously derived from the KS model 
gives the following protocol for simulating entanglement. Alice and Bob share 
the random vectors, $\vec x_1$ and $\vec x_2$,
and perform two projective measurements, $M_A=\{\vec a_1,\vec a_2\}$ 
and $M_B=\{\vec b_1,\vec b_2\}$, respectively.
Bob generates the outcome $\vec b_\beta$ and the indices $n_1=\pm 1$, $n_2=\pm1$ 
according to the probability distribution
$$
\rho(\beta,n_1,n_2|\vec x_1,\vec x_2;M_B)\equiv
\frac{1}{2}\theta\left(n_1\,\vec b_\beta\cdot\vec y_1\right)
\theta\left(n_2\,\vec b_\beta\cdot\vec y_2\right),
$$
He sends $n_i$ to Alice. She generates the outcome $\vec a_\alpha$ according to
the probability distribution defined by Eq.~(\ref{cond2}).

In this model the amount of communication is $2$ bits. It is possible 
to reduce the communication cost by means of the transformation 
$n_1\vec y_1\rightarrow\vec y_1$. Indeed the index $n_1$ becomes
uncorrelated with the other variables and can be eliminated. Thus,
we get a model with a communication cost of $1$ bit, defined by
the conditional probabilities
\be
\label{prob_am}
\rho(\beta,n|\vec x_1,\vec x_2;M_B)\equiv
\theta\left(\vec b_\beta\cdot\vec y_1\right)
\theta\left(n\,\vec b_\beta\cdot\vec y_2\right),
\ee
\vspace{-7mm}
\be
\label{meas_am}
\rho(\alpha|\vec x_1,\vec x_2,n;M_A)=
\theta\left[-\vec a_\alpha\cdot (\vec y_1+n \vec y_2)\right].
\ee
That is, Bob generates the vector $\vec b_\beta$ closest to $\vec y_1$
and sets the discrete index $n$ equal to the sign of $\vec b_\beta\cdot\vec y_2$.
He sends $n$ to Alice. She then generates the outcome $\vec a_\alpha$ that
is closest to $\vec y_1+n \vec y_2$.
Note that this model is very similar to the Toner-Bacon model, with 
the only difference that the unit vectors $\vec x_i$ are replaced by the 
vectors $\vec y_i$, which are a linear combination of $\vec x_i$
[see Eq.~(\ref{vect_y})]. 

This model of entanglement can be put into a more synthetic form that will be 
useful in Sec.~\ref{sec2A}.
Suppose that the measurement outcomes are $\vec a_{\bar\alpha}$ and 
$\vec b_{\bar\beta}$ and the communicated index is $\bar n$;
then it is easy to show by Eq.~(\ref{prob_am}) that
$\vec b_{\bar\beta}\cdot\vec x_{\bar n}\ge\vec b_\beta\cdot\vec x_n$ for any
$\beta\in\{1,2\}$ and $n\in\{1,2\}$. Furthermore from Eq.~(\ref{meas_am}) we have that 
$-\vec a_{\bar\alpha}\cdot\vec x_{\bar n}\ge-\vec a_{\alpha}\cdot\vec x_{\bar n}$
for any $\alpha\in\{1,2\}$. Thus, the algorithm can be reformulated as follows.
Bob evaluates the vectors $\vec b_\beta\in\{\vec b_1,\vec b_2\}$ and
$\vec x_n\in\{\vec x_1,\vec x_2\}$ that maximize $\vec b_\beta\cdot\vec x_n$.
He generates the outcome $\vec b_\beta$ and sends the index $n=\{1,2\}$
to Alice. Alice generates the outcome $\vec a_\alpha\in\{\vec a_1,\vec a_2\}$
that maximizes $-\vec a_\alpha\cdot\vec x_n$. This algorithm will be
generalized to higher dimensions of the Hilbert space in the next section.

\section{Generalizing to higher dimensions}

In Sec.~\ref{sec2A} the three approximate protocols for simulating entanglement 
in higher dimensions are introduced. In Sec.~\ref{sec2B} we will present a
simple method for converting an entanglement protocol into a protocol 
for simulating quantum channels.

\subsection{Entanglement}
\label{sec2A}

Let us consider the following scenario. 
Alice and Bob receive two $N$-dimensional quantum systems in the entangled 
state 
\be\label{bell_state}
|\psi_{AB}\rangle=\frac{1}{\sqrt N}\sum_{k=1}^N
|k\rangle_A|k\rangle_B
\ee
They perform local projective measurements using the
set of orthogonal vectors $\{|\phi_1\rangle,...,|\phi_N\rangle\}\equiv M_A$
and $\{|\psi_1\rangle,...,|\psi_N\rangle\}\equiv M_B$, respectively.
This scenario can be reproduced by local hidden variables augmented
by some amount of communication. 

The gap between the
classical and quantum algorithms in the HM problem implies that
this communication cannot be smaller than $e^{O(\sqrt n)}$ in the
case of bounded error, $n=\log_2 N$ being the number of ebits.
A stronger constraint was given in Ref.~\cite{brassard},
where it was shown that $O(2^n)$ bits of communication are necessary
for an exact simulation. An exact two-way classical protocol with $O(n2^n)$ 
bits of communication on average was reported by Massar et al.~\cite{massar}. 
Unlike the protocols considered in this paper, the protocol in Ref.~\cite{massar}
does not require any local hidden variables. A one-way model of a quantum 
channel for a single qubit was recently reported in Ref.~\cite{montina0}. It
requires an infinite amount of communication for an exact simulation, but
it has the nice property of encoding the communication in a single
real variable, instead of two real variables, which are required for defining a
quantum state. It can be easily converted into a classical protocol of
entanglement. Just as that in Ref.~\cite{massar}, this protocol 
does not need local hidden variables.

In this section we present three approximate protocols for simulating
entanglement that give accurate results for low values of $N$ and require a 
communication of just $n$ bits. They were numerically tested for a family
of measurements. The maximal discrepancy between the best protocol and the 
quantum predictions is less than $1\%$ for $N=3$ and grows sublinearly 
with $n(=\log_2 N)$ in the range numerically studied.
The first protocol is a generalization of the Toner-Bacon model of a
single Bell state, suitably readjusted in Sec.~\ref{tonerSec}. The second 
protocol is a generalization of the alternative model presented in
Sec.~\ref{mymodel}. The last model is similar to the second one, but
with a different shared randomness.

{\bf Protocol $1$}
\begin{enumerate}
\item Bob and Alice share one random vector $|x\rangle$ and a random
basis $\{|y_1\rangle,...,|y_N\rangle\}\equiv Y$.
\item Bob generates the outcome $|\psi_b\rangle\in M_B$ that is closest to 
$|x\rangle$.
\item He sends Alice the index of the vector $|y_m\rangle\in Y$
that is closest to $|\psi_b\rangle$.
\item Alice generates the event $|\phi_a\rangle\in M_A$
that maximizes $\langle\phi_a^*|\left\{|x\rangle\langle x|+
|y_n\rangle\langle y_n|\right\}|\phi_a^*\rangle$,
where $|\phi_a^*\rangle\equiv \sum_k \langle\phi_a|k\rangle|k\rangle$.
\end{enumerate}

Note that this model gives the right marginal distributions for
the outcomes of each party, since the vectors $|x\rangle$ and
$|y_k\rangle$ are generated uniformly in the Hilbert space.
This holds also for the other protocols.
Thus, only the accuracy of the correlations has to be checked. 

{\bf Protocol $2a$}
\begin{enumerate}
\item Bob and Alice share a set $R\equiv\{|x_1\rangle,...,|x_N\rangle\}$
of $N$ random vectors.
\item Bob evaluates the vectors $|\psi_b\rangle\in M_B$ and $|x_n\rangle\in R$
that maximize $|\langle\psi_b|x_n\rangle|^2$.
\item He generates the outcome $|\psi_b\rangle$ and sends $n$ to Alice.
\item Alice generates the event $|\phi_a\rangle\in M_A$
that maximizes $|\langle\phi_a^*|x_n\rangle|^2$.
\end{enumerate}

This protocol is the generalization of the classical model of entanglement
derived in Sec.~\ref{mymodel}.
As shown below, it is more accurate than protocol $1$ for $N\ge10$,
in the sense that the maximal discrepancy (in the set of tested 
measurements) is lower. This is true at least for $N\le32$, which is
the maximal dimension we considered in the numerical simulations.
The last protocol is similar to protocol $2a$, but the vectors $|x_k\rangle$
are orthogonal.

{\bf Protocol $2b$}: the same as protocol $2a$, but Bob and Alice
share a random orthogonal basis.

While this protocol is not exact for $N=2$, it is more accurate than
protocol $2a$ for $N>2$. As shown by the numerical simulations,
the accuracies of protocols $2a$ and $2b$ approach each other 
in the limit of high dimensions.

\begin{figure}
\epsfig{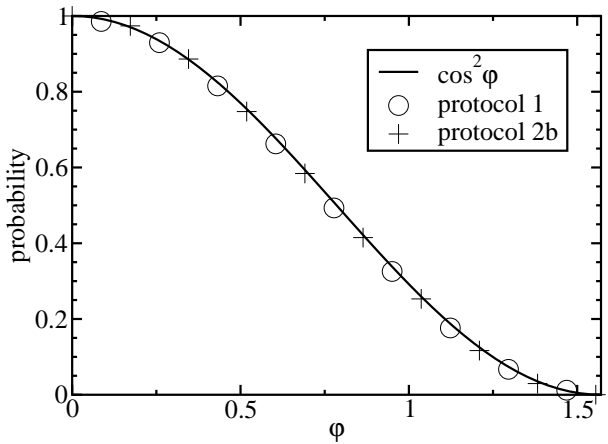}
\caption{Probability of the event $|\phi_1\rangle$ given $|\psi_1\rangle$ as a function 
of $\varphi$ for $N=3$. The solid line is the quantum prediction. The circles and crosses 
are the probabilities obtained from protocols $1$ and $2b$, respectively.}
\label{fig1}
\end{figure}

It is interesting to note that the three protocols share a common 
feature. Kochen and Specker proved that any deterministic hidden variable 
theory equivalent to quantum theory is contextual~\cite{kochen}. Although 
our protocols are not completely
equivalent to quantum theory, they satisfy this general constraint, 
that is, the probability of a joint event $(|\phi_\alpha\rangle,
|\psi_\beta\rangle)$ depends on the whole set of orthogonal vectors 
$M_A$ and $M_B$. Indeed, all the vectors in the sets are involved
in the maximization procedures used by the protocols.  
It is interesting to note that a similar procedure of maximization 
was used in an approximate hidden-variable model of qutrit 
reported in Ref.~\cite{rudolph}. The Kochen-Specker 
theorem was recently generalized to probabilistic theories in 
Ref.~\cite{montina}.

Since these protocols are contextual and approximate, we can expect
that the probability distribution of two events $|\psi_\beta\rangle$
and $|\phi_\alpha\rangle$ again depends on all the vectors in the sets
$M_A$ and $M_B$. For the sake of simplicity, here we will consider 
the set of one-parameter measurements 
\be
\label{measu}
\begin{array}{l}
|\phi_1\rangle=\cos\varphi |\psi_1\rangle-\sin\varphi |\psi_2\rangle, \\
|\phi_2\rangle=\sin\varphi |\psi_1\rangle+\cos\varphi |\psi_2\rangle, \\
|\phi_k\rangle=|\psi_k\rangle\; \mathrm{for  }\; k=3,...,N,
\end{array}
\ee
that is, the vectors $|\phi_k\rangle$ are set equal to $|\psi_k\rangle$ for
$k>2$. The results do not change qualitatively with a different choice of 
the measurements.

The quantum probabilities of the joint events $|\phi_b\rangle$ and $|\psi_a\rangle$
for $a\ne1,2$ or $b\ne1,2$ are obviously equal to $\delta_{a,b}$. 
The probabilities $P(a,b|\varphi)$ of the outcomes $a=1,2$ and $b=1,2$, given
$\varphi$, are 
\bey
P(1,1|\varphi)=P(2,2|\varphi)=\frac{1}{N}\cos^2\varphi  \\
P(1,2|\varphi)=P(2,1|\varphi)=\frac{1}{N}\sin^2\varphi.
\eey

Note that the measurement with $\varphi=\varphi_0$ is equivalent to the measurement
with $\varphi=\frac{\pi}{2}-\varphi_0$ and $|\psi_1\rangle$ and $|\psi_2\rangle$
swapped. Furthermore, $\sum_{k,l} P(k,l|\varphi)=1$. Thus, it is sufficient to
evaluate the discrepancy between the model and quantum theory for outcomes 
$a=b=1$ and $\varphi=[0,\pi/2]$. This discrepancy contains the full information about
the discrepancy of any other event with constraint~(\ref{measu}).

In Fig.~\ref{fig1}, we report the probability of an event $|\phi_1\rangle$, given
$|\psi_1\rangle$, as a function of $\varphi$ for $N=3$. 
The solid line is the quantum prediction $\cos^2\varphi$. The circles
and crosses are the Monte Carlo results, generated by the protocols $1$ and $2b$.
The maximal discrepancy between protocol $2b$ and $\cos^2\varphi$ is very
small, less than $0.01$, and is barely perceptible in the figure. 
Protocol $2a$ gives the worst results with a
maximal discrepancy about $0.025$. Its data are not reported in the figure.
The maximal discrepancy for protocol $1$ is about $0.014$. The numerical 
simulations were performed with $10^7$ Monte Carlo executions, which make the 
statistical error inappreciable in the figure. 

\begin{figure}[h]
\epsfig{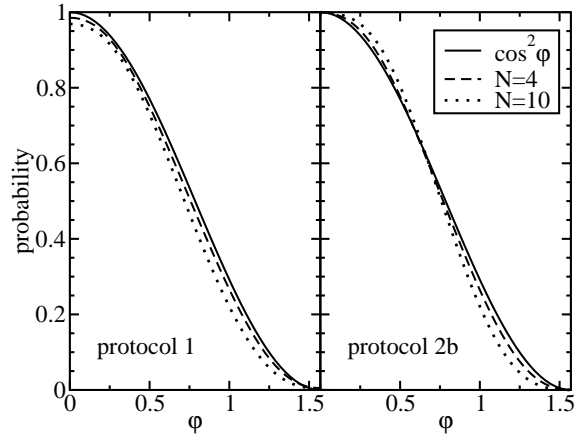}
\caption{Same as Fig.~\ref{fig1} with $N=4$ (dashed line) and $N=10$ (dotted line).}
\label{fig2}
\end{figure}

The same simulations were executed for higher dimensions of the Hilbert space;
the results for $N=4$ (two qubits) and $N=10$ are plotted in Fig.~\ref{fig2}. 
The probabilities generated by protocol $1$($2b$) are reported at the left-hand 
(right-hand) 
side. It is interesting to note that protocol $2b$ gives the right probability 
for $\varphi=0,\pi/2$, that is, when $|\phi_1\rangle$ and $|\psi_1\rangle$ are 
parallel or orthogonal. However, a small discrepancy was noted for measurements 
that do not satisfy constraint~(\ref{measu}). 

\begin{figure}
\epsfig{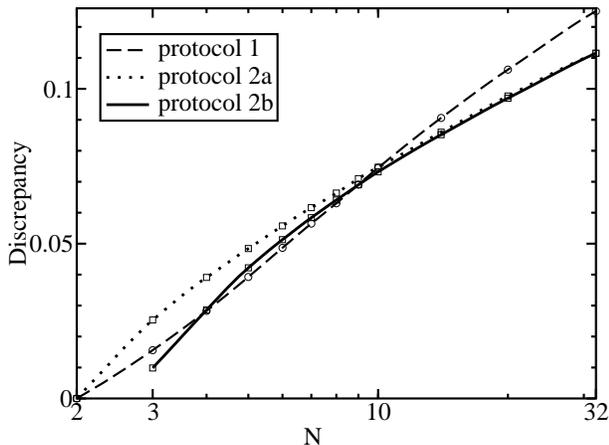}
\caption{Maximum discrepancy between the quantum prediction and the classical
protocols as a function of $N$. The data are interpolated by lines, plotted as a 
guide of eyes.}
\label{fig3}
\end{figure}

The maximum discrepancy of the three protocols is reported in Fig.~\ref{fig3} 
as a function of $N$. The horizontal axis is in logarithmic scale. For large
values of $N$, protocols $2a$ and $2b$ have the same discrepancy. This is
because in protocol $1$ the vectors $|x_k\rangle$, which are generated 
randomly, have a very high probability of being almost orthogonal to each 
other for $N\gg1$.
This fact makes protocol $2a$ almost indistinguishable from protocol $2b$
in high dimensions.

\subsection{Quantum channel}
\label{sec2B}

In Ref.~\cite{cerf} it was shown that any protocol that simulates entanglement
of two qubits can be converted into a classical model of a quantum channel for
a single qubit. A very simple method for converting a general classical model of 
entanglement into a classical model of a quantum channel is as follows. 
In the entanglement model, Bob and Alice share a set $R$ of random vectors.
This means that
they have a common list of noise realizations $R_k$, with $k=1,2,3,...$.
They start from $k=1$ and at each execution of the Monte Carlo 
simulation they read the next element of the list. 
It is possible to convert the protocol into a protocol for simulating
quantum channels by increasing the communication by $n$ bits on average.
Suppose that Bob receives the quantum state $|\psi\rangle$. He selects
a measurement $M_B$=\{$|\psi_1\rangle$,...,$|\psi_N\rangle$\} so that
$|\psi_1\rangle=|\psi\rangle$. Using the protocol for simulating
entanglement and the first noise realization in the shared list, Bob generates
a vector $|\psi_b\rangle$. If $|\psi_b\rangle\ne|\psi\rangle$, Bob 
interrupts the protocol for simulating entanglement and
reads the next realization of the noise in the shared list; he repeats the 
procedure until $|\psi_b\rangle=|\psi\rangle$.  He then executes the communication 
procedure as established by the entanglement protocol. Furthermore he sends the 
number of noise realizations that Alice has to skip. This additional information 
is equal to $\log_2 N=n$ on average.

Thus, using this strategy, it is possible to convert the three models 
of entanglement into protocols for simulating quantum channels. This
conversion requires doubling the amount of communication on average.

\section{Conclusion}

In this paper we have presented three approximate one-way communication protocols 
for simulating the outcomes of local measurements, performed on bipartite 
entangled states. These protocols use an amount of communication equal to the number 
$n$ of ebits. We have seen that they can be converted into approximate protocols for 
simulating quantum communication channels. Approximate models like 
these can be useful for detecting if a quantum communication algorithm 
can be efficiently simulated by some classical algorithm. This is the
case when
a quantum communication algorithm uses states and measurements that an
approximate model of a quantum channel can efficiently simulate
with zero or bounded error in any dimension.

The results reported in this paper can be improved in different ways.
It is interesting to note that
the protocols $2a$ and $2b$ have the same general structure, but 
different shared noises. The general structure is as follows.
Bob and Alice share a set $R\equiv\{|x_1\rangle,...,|x_N\rangle\}$
of $N$ random vectors with probability distribution 
$\rho(|x_1\rangle,...,|x_N\rangle)$.
Bob evaluates the vectors $|\psi_b\rangle\in M_B$ and $|x_n\rangle\in R$
that maximize the function 
\be\label{gen_model_1}
F_1(b,n)\equiv|\langle\psi_b|x_n\rangle|^2.
\ee
He generates the outcome $|\psi_b\rangle$ and sends $n$ to Alice.
Alice generates the event $|\phi_a\rangle\in M_A$
that maximizes the function 
\be\label{gen_model_2}
F_2(a)\equiv|\langle\phi_a^*|x_n\rangle|^2.
\ee
We have seen that a suitable choice of the 
noise distribution can considerably reduce the error. Indeed, in
the three-dimensional case a change of noise dropped the error
from $2.5\%$ (protocol $2a$) to less than $1\%$ (protocol $2b$). 

Thus, protocols $2a$ and $2b$ can be improved by evaluating the 
optimal probability distribution $\rho(|x_1\rangle,...,|x_N\rangle)$
that minimizes the maximal or average error over a set of measurements.
Of course, the error cannot be reduced to zero if the whole set of
measurements is considered, since this would require an exponential
amount of communication. However, another possible improvement
can be reached by augmenting the communication cost, namely 
$\log_2 N$. This is achieved by increasing the number $N$ of random 
vectors $|x_k\rangle$. We can expect that this strategy and the 
optimal choice of $\rho$ will enhance the accuracy of the model, as 
defined through the functions in Eqs.~(\ref{gen_model_1},\ref{gen_model_2}).
For a sufficiently large amount of communication, one can hope
to make the error bounded. Notice that, as said in the introduction, the 
{\it HM} problem establishes the lower bound $2^{(\Omega(\sqrt{n})}$ for 
the communication cost, whereas the best known protocols require an amount 
of communication that scales as $n2^n$~\cite{buhrman,massar}. At present 
it is not clear if such protocols are also optimal. This open problem can 
be investigated by studying the class of protocols introduced in this paper.
In conclusion, our models
can give some indication for finding optimal one-way communication protocols 
that classically simulate quantum channels and entanglement. Furthermore,
they can be used for testing the efficiency of a quantum communication
protocol versus classical protocols.

\section*{Acknowledgments}
The author acknowledges useful discussions with Rob Spekkens and
Erik Schnetter.
Research at Perimeter Institute for Theoretical Physics is
supported in part by the Government of Canada through NSERC
and by the Province of Ontario through MRI.

\end{document}